\documentclass[12pt]{article}

\usepackage{amsmath,amsfonts,amssymb,cite,graphicx}

\def\bea{\begin{eqnarray}}
\def\eea{\end{eqnarray}}
\def\be{\begin{equation}}
\def\ee{\end{equation}}
\def\ba{\begin{array}}
\def\ea{\end{array}}
\def\nn{\nonumber}

\newcommand{\dm}{\hat{d}}

\begin{document}

\setlength\arraycolsep{2pt}

\renewcommand{\theequation}{\arabic{section}.\arabic{equation}}
\setcounter{page}{1}

\setlength\arraycolsep{2pt}

\begin{titlepage}

\rightline{\footnotesize{DESY 08-198}} \vspace{-0.2cm}
\rightline{\footnotesize{CERN-PH-TH/2008-247}} \vspace{-0.2cm}

\begin{center}

\vskip 0.4 cm

{\LARGE  \bf Constructing de Sitter vacua in no-scale \\[1mm]
string models without uplifting}

\vskip 0.7cm

{\large 
Laura Covi$^{a}$, Marta Gomez-Reino$^{b}$, Christian Gross$^{c}$, \\[1mm]
Gonzalo A. Palma$^{d}$, Claudio A. Scrucca$^{e}$
}

\vskip 0.5cm

{\it
$^{a}$Theory Group, Deutsches Elektronen-Synchrotron DESY,\\ 
D-22603 Hamburg, Germany\\
$^{b}$Theory Division, Physics Department, CERN, \\
CH-1211 Geneva 23, Switzerland\\
$^{c}$II. Institut f\"ur Theoretische Physik, Universit\"at Hamburg, \\
D-22761 Hamburg, Germany\\
$^{d}$Lorentz Institute for Theoretical Physics, Leiden University, \mbox{NL-2333 CA Leiden, The Netherlands}\\
$^{e}$Inst. de Th. des Ph\'en. Phys.,
Ecole Polytechnique F\'ed\'erale de Lausanne, \mbox{CH-1015 Lausanne, Switzerland}\\
}

\vskip 0.8cm

\end{center}

\begin{abstract}

We develop a method for constructing metastable de Sitter vacua in $\mathcal N=1$ supergravity models 
describing the no-scale volume moduli sector of Calabi-Yau string compactifications. We consider both heterotic 
and orientifold models. Our main guideline is the necessary condition for the existence of metastable vacua 
coming from the Goldstino multiplet, which constrains the allowed scalar geometries and supersymmetry-breaking
directions. In the simplest non-trivial case where the volume is controlled by two moduli, this condition simplifies 
and turns out to be fully characterised by the intersection numbers of the Calabi-Yau manifold. 
We analyse this case in detail and show that once the metastability condition is satisfied it is possible to 
reconstruct in a systematic way the local form of the superpotential that is needed to stabilise all the fields. 
We apply then this procedure to construct some examples of models where the superpotential
takes a realistic form allowed by flux backgrounds and gaugino condensation effects, for which a viable vacuum 
arises without the need of invoking corrections to the K\"ahler potential breaking the no-scale property or uplifting 
terms. We finally discuss the prospects of constructing potentially realistic models along these lines.

\end{abstract}

\end{titlepage}

\newpage

\section{Introduction} \setcounter{equation}{0}

Current cosmological observations convincingly suggest that our universe is undergoing an accelerated expansion. 
The simplest model accounting for this result involves backgrounds with a tiny positive cosmological 
constant. This has lead in the past years to a lot of activity in the search of de Sitter (dS) vacua in the four-dimensional 
low-energy effective supergravity description of string theory compactifications. It is now well understood that 
effects like gaugino condensation and background fluxes can induce terms in the effective 
superpotential that allow to stabilise many or even all of the moduli fields. However, this generically leads to a 
supersymmetric ground state which is either anti-de Sitter (AdS) or Minkowski space, and it is surprisingly difficult 
to obtain non-supersymmetric dS  vacua \cite{Giddings:2001yu, Saltman:2004sn, 
Balasubramanian:2005zx, lust}. One generic way of overcoming this difficulty is to start from a setting leading to 
an AdS vacuum and add to it some additional sources of hard supersymmetry breaking, like anti-D3 branes 
\cite{Kachru:2003aw} or other localised sources \cite{Silverstein:2007ac,Haque:2008jz}, to uplift the vacuum 
energy. However, the addition of such sources does not admit a transparent effective supergravity description, 
and refinements of this scenario have been considered where the uplifting sector breaks supersymmetry softly
and contains additional light degrees of freedom \cite{Burgess:2003ic, choi,Lebedev:2006qq, vz, carlos, jeong, Becker:2004gw, Saueressig:2005es, Achucarro:2007qa}.
Alternatively, one may achieve dS vacua in a more genuine way thanks to leading perturbative or non-perturbative 
corrections to the K\"ahler potential \cite{Becker:2002nn,Balasubramanian:2004uy,Parameswaran:2006jh,Palti:2008mg,Berg:2007wt}. 
In that case, however, one has to make sure that higher-order subleading corrections are under control. 

Despite of the success of the above approaches in producing viable vacua, it would be desirable to have models where metastability is granted from the onset, without the need to incur into either subleading corrections 
or an additional uplifting sector for help. Ideally, one may want to achieve this within the sector of the moduli 
fields. The simplest option could be to use just the dilaton, which universally spans the coset space 
$SU(1,1)/U(1)$, but this has been excluded unless uncontrollably large corrections arise for the geometry 
\cite{Casas:1996zi,Brustein:2004xn,GomezReino:2006dk}. Another interesting possibility could be to use 
only the volume moduli (also called K\"ahler moduli), which have the universal characteristics of spanning a 
scalar manifold with a no-scale property. Interestingly, no explicit example is known so far where a viable vacuum 
is produced without invoking corrections to the K\"ahler potential breaking its no-scale structure. 
In the simplest cases where the moduli space is a coset manifold with covariantly-constant curvature, 
like in the case of one modulus or more generically for $n$ moduli in orbifold limits of Calabi-Yau (CY) compactifications, 
it has been proved in \cite{GomezReino:2006dk, GomezReino:2006wv} (see also \cite{Brustein:2004xn}) that dS 
vacua are in fact unavoidably unstable, because one of the scalar partners of the Goldstino always has a semi-negative 
mass-squared, for any superpotential. It was however shown later in \cite{Covi:2008ea} (see also \cite{Covi:2008cn}) 
that this no-go theorem can be evaded when the moduli span a less constrained space, 
like for smooth CY compactifications. 
One of the main results deduced in~\cite{Covi:2008ea}, following the line of reasoning of \cite{GomezReino:2006dk,
GomezReino:2006wv}, is a necessary condition on the K\"ahler geometry of the moduli space 
for a metastable dS vacuum to possibly arise. This condition depends on the intersection numbers $d_{ijk}$ and thus 
restricts the type of CY manifold that can be used. Furthermore, it also constrains the direction in field
space along which supersymmetry is allowed to be broken, and thus implicitly restricts the form of the superpotential as well. 

The aim of this paper is to analyse in more detail such models, and to study how to determine a 
superpotential which allows for metastable de Sitter vacua for a given choice of CY manifold.
We shall focus on the simplest non-trivial class of models involving two volume moduli, for which the metastability
condition simplifies and can be made more explicit, but we believe that the situation for models with more volume 
moduli should be qualitatively similar. We will then look for a systematic procedure to reconstruct the required form 
of the superpotential that is needed to achieve stabilisation of all the moduli, once the metastability condition on 
the K\"ahler geometry is satisfied.

The paper is organised as follows. In Section \ref{sec: metastab} we briefly review the results of 
refs. \cite{Covi:2008ea, GomezReino:2006dk, GomezReino:2006wv} regarding the metastability of 
supersymmetry-breaking vacua and their implications. In Section \ref{sec: LVM} we apply these results 
to the more particular case of CY string models with two volume moduli, and deduce which type 
of models can possibly allow viable vacua. In Section \ref{sec: de Sitter - two mod} we further analyse 
those models satisfying the metastability condition, and describe a procedure to determine the type of 
superpotential that is required to actually get a metastable dS vacuum. In Section \ref{sec: string models} 
we provide explicit examples of string models with a volume moduli sector satisfying all these requirements
and admitting a metastable dS vacuum. Finally, in Section \ref{sec: conc} we make some concluding remarks.

\section{Metastability in supergravity} \setcounter{equation}{0}  \label{sec: metastab}

Let us start by reviewing the analysis of the stability of non-supersymmetric vacua with non-negative 
cosmological constant in $\mathcal N = 1$ supergravity models, following 
refs.~\cite{GomezReino:2006dk,GomezReino:2006wv} and \cite{Covi:2008ea,Covi:2008cn}.\footnote{See 
\cite{GomezReino:2008bi} for a similar analysis in the context of $\mathcal N=2$ supergravity with 
only hypermultiplets.}
We assume here that vector multiplets play a negligible role in the dynamics of supersymmetry breaking 
and focus thus on theories involving only chiral multiplets.\footnote{See \cite{GomezReino:2007qi} for 
a study of the effects of vector multiplets.}

Recall first that the most general two-derivative Lagrangian for a supergravity theory with $n$ chiral 
superfields is entirely determined by the function $G=K+\ln|W|^2$, which depends on the chiral 
superfields $\Phi^i$ and their conjugates $\bar \Phi^{\bar \imath}$ through a real K\"ahler potential 
$K$ and a holomorphic superpotential $W$.\footnote{We set $M_{Pl} = 1$ and denote derivatives 
with respect to $\phi^i$ and $\bar \phi^j$ by lower indices $i$ and $\bar \jmath$.}
The scalar fields span a K\"ahler manifold with a metric given by $g_{i \bar \jmath} = K_{i \bar \jmath}$,
for which the only non-vanishing components of the Christoffel connection and Riemann tensor are 
$\Gamma_{ij}^k = g^{k \bar l} K_{i j \bar l}$ (and its conjugate), and 
$R_{i \bar \jmath m \bar n} = K_{i \bar \jmath m \bar n} - K_{i m \bar l} g^{\bar l k}K_{k \bar \jmath \bar n}$ 
(and permutations). The chiral auxiliary fields are fixed by their equations of motion to be $F^i=m_{3/2}G^i$, 
with a scale set by the gravitino mass $m_{3/2}=e^{G/2}$. Whenever $F^i\neq 0$ at the vacuum, supersymmetry 
is spontaneously broken and the direction $G^i$ in the space of chiral fermions defines the Goldstino fermion 
which is absorbed by the gravitino in the process of supersymmetry breaking.
We shall describe this direction also in the scalar field space by the 
unit vector
\be
f_i = \frac {G_i}{\sqrt{G^k G_k}} \,.
\ee
Moreover, we will parametrise the cosmological constant in terms of 
the gravitino mass through the dimensionless quantity
\be
\gamma = \frac {V}{3 m_{3/2}^2} \,.
\ee

The scalar fields have a kinetic term controlled by the K\"ahler metric 
$g_{i \bar \jmath}$, which is thus assumed to be positive-definite, 
and a potential $V$ that takes the following simple form:
\be
V = e^{G} \big(G^{i} G_{i} - 3  \big)\,.
\label{s2:F-term-G} 
\ee
Supersymmetry-breaking metastable vacua with non-negative cosmological 
constant are associated to local minima of the potential at which 
$G^i \neq 0$ and $V \ge 0$. The $n$ complex stationarity condition are derived 
by computing $V_i = \nabla_i V$ and read:
\be
V_i = e^{{G}} \left({G}_{i} + {G}^{k} \nabla_{i} {G}_{k}\right) + G_i V =0\, .
\label{stationary1}
\ee
The $2n$ dimensional mass matrix for scalar fluctuations around such a vacuum takes the form
\begin{eqnarray}
\label{M}
M^2 = \left(\begin{matrix}
V_{i \bar \jmath} & V_{i j} \\
V_{\bar \imath \bar \jmath} & V_{\bar \imath j}
\end{matrix}\right) \,,
\end{eqnarray}
in terms of the second derivatives of the potential $V_{i \bar \jmath} = \nabla_i \nabla_{\bar \jmath} V$ and
$V_{i j} = \nabla_i \nabla_j V$, which can also be computed using covariant derivatives since the extra 
connection terms vanish by the stationarity conditions, and read:
\begin{eqnarray}
V_{i \bar \jmath} &=&  e^{G} \left(G_{i \bar \jmath}  +  \nabla_i G_k \nabla_{\bar \jmath}  G^k
-  R_{i \bar \jmath m \bar n} \, G^m  G^{\bar n} \right) + \left(G_{i \bar \jmath}  -  G_i G_{\bar \jmath}\right) V \,,
\label{mass1} \\[1mm]
V_{i j} &=& e^{G} \left(2 \nabla_{(i} G_{j)}+  G^k \nabla_{(i} \nabla_{j)} G_k \right)
+ \left(\nabla_{(i} G_{j)} -  G_i G_j \right) V \,. 
\label{mass2}
\end{eqnarray}
The metastability condition is then the requirement that the $2n$-dimensional 
mass matrix (\ref{M}) should be positive definite. 

\subsection{Necessary condition for metastability}  \label{stability analysis} 

As discussed in detail in \cite{Covi:2008ea, Covi:2008cn} it is clear that for a fixed K\"ahler 
potential $K$, most of the eigenvalues of $M^2$ can be made positive and arbitrarily 
large by suitably tuning the superpotential $W$. The only restriction comes from the fact 
that the projection of $V_{i \bar \jmath}$ along the Goldstino direction $f^i$ is actually 
constrained by the stationarity conditions (\ref{stationary1}), which imply
$\nabla_i G_j f^j = - (1 + 3\gamma)f_i$, and therefore cannot be adjusted so easily. 
As a consequence of this fact, in order to study metastability it is sufficient to study the projection of the diagonal block 
$V_{i \bar \jmath} $ of the mass matrix along the Goldstino direction. 
This projection defines a mass scale $m$ which is related to the 
masses of the two sGoldstinos and is given by
\be
m^{2}  \equiv  V_{i \bar \jmath} \, f^i f^{\bar \jmath} \, .  
\label{def m-tilde}
\ee
A necessary condition for the mass matrix (\ref{M}) to be positive-definite is that 
$m^2 > 0$. One can then compute this quantity more explicitly and derive 
a necessary condition for metastability of the vacuum. By using eqs.~(\ref{stationary1}) 
and (\ref{mass1}), one finds: 
\be
m^2 = \left[ 3(1+\gamma) \hat \sigma(f^{i}) - 2 \gamma \right] m_{3/2}^2 \,,
\label{m1}
\ee
where\footnote{We use the same notation as in \cite{Covi:2008cn}
for this quantity, the hat being introduced to distinguish it from the quantity 
$\sigma$ defined in \cite{Covi:2008ea}, which has a different normalisation.}
\be 
\label{sigma}
\hat \sigma (f^i) \equiv \frac23 - R_{i \bar \jmath m \bar n} \, f^{i}  f^{\bar \jmath} f^{m} \! f^{\bar n}\,  .
\ee
The condition $m^2 > 0$ implies then the constraint
\be
\hat \sigma(f^{i}) > \frac 23 \frac{\gamma}{1+\gamma} \,.
\label{neccond}
\ee
Observe that the quantity $R_{i \bar \jmath m \bar n} \, f^{i}  f^{\bar \jmath} f^{m} \! f^{\bar n}$ in eq.~(\ref{sigma}) 
corresponds to the holomorphic sectional curvature along the Goldstino vector $f^i$ and therefore 
eq.~(\ref{neccond}) is a restriction on the allowed scalar geometries and supersymmetry breaking 
directions.

Notice that for a fixed $K$ and arbitrary $W$, the direction $f^i$ can be varied while 
keeping the metric and the Riemann tensor fixed. One can then look for the preferred 
direction $f_0^i$ that maximises $m^2$ with value $m^2_0$. 
If $m^2_0 < 0$, then one of the sGoldstinos is unavoidably tachyonic, and the vacuum is 
unstable. If instead $m^2_0 >0$, then the sGoldstinos can be kept non-tachyonic 
by choosing $W$ such that $f^i$ is close-enough to $f_0^i$. As already mentioned, the rest of 
the scalars can always be given a positive square mass by further tuning $W$.

\subsection{The sGoldstino mass}  \label{Section: sGoldstino}

As noted above, $m^2$ is related to the square masses of the sGoldstinos, but in general it does
not exactly coincide with them, since $f^i$ is in general not an eigenvector of the full mass matrix (\ref{M}).
We will now show that the preferred direction $f_0^i$ is instead automatically an eigenvector of the 
diagonal blocks of (\ref{M}), and the corresponding mass $m_0^2$ is then more directly related to 
their mass eigenvalues. More precisely, when the off-diagonal block of (\ref{M}) vanishes one has two 
degenerate sGoldstinos with square masses given by $m_0^2$, whereas when the off-diagonal 
block does not vanish these two masses split.

To prove this statement, let us determine implicitly the direction $f_0^i$ for which $m^2$ reaches 
its maximum value $m_0^2$. To do this, we vary the unit vector $f^i$ while keeping 
the vacuum expectation values (vevs) of the chiral fields fixed, and try to maximise $\hat \sigma(f^i)$. 
Enforcing the constraint $f^i f_i = 1$ with the help of a Lagrange multiplier $\xi$, we are then led to extremise 
the following functional:
\be
F(f^i,\xi) = \hat \sigma (f^i) + \xi \big(g_{i \bar \jmath} f^i f^{\bar \jmath} - 1\big) \,.
\ee
Stationarity with respect to $f^i$ implies the relation 
$f_{0i} = 2 \xi_0^{-1} R_{i \bar \jmath m \bar n} f_0^{\bar \jmath} f_0^{m} f_0^{\bar n}$,
which implicitly defines the values of $f_0^i$ in terms of $\xi_0$. Plugging this result back into 
the constraint $f_0^i f_{0i} = 1$, which follows from stationarity with respect to $\xi$, 
determines then $\xi_0 = 2 R_{i \bar \jmath m \bar n} f_0^i f_0^{\bar \jmath} f_0^{m} f_0^{\bar n}$.  
Putting everything together, one finally finds the following relation implicitly determining $f_0^i$:
\be
f_{0i} =  \frac{R_{i \bar \jmath m \bar n} f_0^{\bar \jmath} f_0^{m} f_0^{\bar n}}
{R_{p \bar q r \bar s} f_0^p f_0^{\bar q} f_0^r f_0^{\bar s}} \,.   \label{f-maximal}
\ee
Using this relation and the stationarity condition (\ref{stationary1}), one can now easily 
verify that $f_{0i}$ is indeed an eigenvector of the matrix $V_i^j$ with 
eigenvalue $m_0^2$:
\be
V_i^j f_{0j} = m_0^2 f_{0i} \,.
\ee

\section{String models with two moduli} \setcounter{equation}{0}  \label{sec: LVM}

In this section we will consider more specifically a class of supergravity models arising 
from the volume moduli sector of CY string compactifications in the low-energy 
and large-volume limit. We assume that the dilaton and complex structure 
moduli do not play any relevant role. We will moreover assume that there are only two
volume moduli, or that possible additional ones do not play any relevant role either. 
We will not address in this paper the circumstances under which such a situation can 
be honestly achieved by making the additional moduli heavy and integrating them 
out.\footnote{See refs.~\cite{Choi:2004sx,deAlwis:2005tf,Achucarro:2008fk,Gallego:2008qi}
for work in this direction.}
Our aim is thus mainly to exhibit the behaviour of a set of two volume moduli with a 
no-scale K\"ahler potential.

\subsection{General properties}

Let us start by recalling a few general properties of these types of models, which 
actually hold true for an arbitrary number of volume moduli. A first important property
is that at leading order in the perturbative and low-energy expansions the effective 
K\"ahler potential satisfies the no-sale property
\be
K^{i}K_{i} = 3 \,.
\label{nosc}
\ee
A second property is that $K$ depends only on $\Phi^i + \bar \Phi^i$, i.e. each 
field enjoys an independent shift symmetry, under which $\delta \phi^i = i \lambda$. 
This allows to drop any distinction between holomorphic and antiholomorphic 
indices in quantities deduced from $K$. Actually, it turns out that there exists 
a special coordinate frame in which $e^{-K}$ is a homogeneous function of degree 
$3$ in the fields $\Phi^i + \bar \Phi^i$. One then has:
\be
- \big(\Phi^i + \bar \Phi^i \big) K_i = 3 \,. 
\label{hom}
\ee
Taking a derivative of this relation it then also follows that $K^i = - \big(\Phi^{i} + \bar \Phi^{i} \big)$. 
This equation, together with (\ref{hom}), implies the no-scale property (\ref{nosc}), and is thus stronger 
than it.

In the light of the above properties, it proves convenient to introduce the unit vector defined by the 
derivatives of the K\"ahler potential:
\be
k_i = \frac 1{\sqrt{3}} K_i \,.
\ee
It was shown in \cite{Covi:2008ea} that as a result of the no-scale property the function $\hat \sigma$
controlling the mass $m^2$ vanishes along this direction, for any value of the fields:
\be
\label{shat}
\hat \sigma(k^i) = 0 \,.
\ee 
As thoroughly discussed in~\cite{Covi:2008ea}, this result allows to study the metastability condition 
by analysing the behaviour of $\hat \sigma(f^i)$ in the vicinity of $f^i = k^i$. In this analysis, a special 
role is played by the subspace orthogonal to $k^i$, which is spanned by a basis of $n-1$ complex 
unit vectors orthogonal to $k^i$.

\subsection{Models with two moduli}

The general problem of determining whether a dS vacuum may arise in the models under consideration 
is still quite complicated, even in the light of the restrictions (\ref{nosc}),  (\ref{hom}) and (\ref{shat}). 
However, one can fully characterise the metastability condition for two-moduli models. 
In this case, the field space is of complex dimension $2$ and can be conveniently parametrised with
a basis of two unit vectors: $k^i$ and a vector $n^i$ perpendicular to it:
\be
k^i n_i = 0 \,.
\ee
This condition defines $n^i$ uniquely, up to an overall phase, in terms of the components
of $k^i$ and the elements of the metric and its inverse. Denoting by $\det g$ the determinant of the metric, 
one easily finds:
\be
(n_1 , n_2) = \sqrt{\det g} \, (k^2, - k^1) \,, \qquad  (n^1 , n^2) = \frac{1}{\sqrt{\det g}} (k_2, - k_1)\,.
\ee
Since the space perpendicular to $k^i$ is one-dimensional, it coincides with the space parallel to $n^i$,
and the projection operator $P^{i j}$ onto such a subspace is simply given by
\be
P^{ij} = g^{ij} - k^i k^j = n^i n^j \,.
\ee
We may now decompose the unit vector $f^{i}$ defining the Goldstino direction in terms of the two 
orthogonal vectors $n^i$ and $k^i$. Up to an overall phase, that we shall not display explicitly, 
we can parametrise the result in terms of an angle $\chi$ and a relative phase $\delta$, and write:
\begin{align}
f^i &= \sin \chi \, k^i +  e^{i \delta} \cos \chi \, n^i  \,, \qquad 
f_i = \sin \chi \, k_i +  e^{-i \delta} \cos \chi \, n_i  \,, \nn \\
f^{\bar \imath} & = \sin \chi \, k^i +  e^{-i \delta} \cos \chi \, n^i \,, \hspace{17pt}
f_{\bar \imath} = \sin \chi \, k_i +  e^{i \delta} \cos \chi \, n_i \,.
\label{f-2-mod}
\end{align}

To proceed further and be more explicit, we need now to distinguish between the two classes 
of heterotic and orientifold models. In ref.~\cite{Covi:2008ea} it was found that in both cases
the possibility of achieving a metastable dS vacuum is linked to the sign of the discriminant
$\Delta$ of the cubic polynomial defined by the intersection numbers $d_{ijk}$, after scaling out one 
variable, and reads
\begin{eqnarray}\label{del}
\Delta = - 27 \Big( d_{111}^2 d_{222}^2 - 3\, d_{112}^2 d_{122}^2 
+ 4\, d_{111} d_{122}^3 + 4\, d_{112}^3 d_{222} 
- 6\, d_{111} d_{112} d_{122} d_{222} \Big)\,. 
\end{eqnarray}
If $\Delta < 0$ the heterotic version can potentially admit dS vacua but not the orientifold one.
Viceversa, if $\Delta > 0$ the orientifold version can but the heterotic cannot. In what follows we compute 
$\hat \sigma$ explicitly in terms of $\chi$ and $\delta$ parameterising $f^i$ for both of these cases.

\subsection{Heterotic models}

In heterotic models, the effective K\"ahler potential takes the following simple form in 
the large volume limit:\footnote{The discussion of this section is also valid for certain classes of orientifold compactifications where the K\"ahler potential exhibits the same form (\ref{K-heterotic}). An example of this are compactifications of type IIB with O5/O9-orientifold planes 
\cite{Grimm:2004uq}.}
\be
K = - \log \mathcal{V} \,, \qquad \mathcal{V} = 
\frac 43 \, d_{i j k}\, t^i t^j t^k \,. \label{K-heterotic}
\ee
In this expression, $d_{ijk}$ denotes the intersection numbers of the CY 
manifold and $t^i$ are the volume moduli. In this case, the $t^i$ can be promoted
in a simple way to (scalar components of) chiral superfields, by setting $t^i= (T^i + \bar T^i)/2$.

From the form of the K\"ahler potential (\ref{K-heterotic}) it follows that 
$K^i = - (T^i + \bar T^i)$ and $K_i = - 1/2\, e^K d_{i m n} K^m K^n$. 
The metric and the Riemann tensor are then given by (see \cite{Covi:2008ea} for more details)
\begin{align}
g_{i j} & =  e^K d_{i j n} K^n + K_i K_j  \label{gij} \,,  \\
R_{i j m n} & = g_{i j} g_{m n} + g_{i n} g_{m j} - e^{2K} d_{i m p} g^{p q} d_{q j n} \label{Rijmn} \,. 
\end{align}
Using this expression, as well as (\ref{f-2-mod}) it is then possible to rewrite $\hat \sigma (f^i)$ 
in the form $\hat \sigma_{\mathcal{H}} (f^i) = - 2 \hat s^i \hat s_i + \hat \omega$, where
\bea
\hat s^i &=& n^i \bigg[ \frac{2}{\sqrt{3}} \tan \chi \, \cos \delta  
- \frac 12 e^{K} d_{p q r} n^p n^q n^r \bigg] \cos^2 \chi \,, \label{s-heterotic} \\
\hat \omega &=& \bigg[\frac 32 \Big( e^K d_{p q r} n^p n^q n^r \Big)^2 - 1 \bigg] \cos^4 \chi \label{omega-heterotic} \,.
\eea
On the other hand, it was shown in \cite{Covi:2008ea} that
\begin{eqnarray} \label{eq:defa} 
\frac 32 \Big( e^K d_{p q r} n^p n^q n^r \Big)^2 - 1 = a_{\mathcal{H}}  \,,
\end{eqnarray}
where
\be
a_{\mathcal{H}}  \equiv - \frac{\Delta}{24} \, \frac {e^{4 K}}{(\det g)^3} \ge -1\,.
\ee
Putting all of these results back into eqs.~(\ref{s-heterotic})-(\ref{omega-heterotic}), 
and introducing the sign $s_{\mathcal{H}} = \mathrm{sign}(d_{p q r} n^p n^q n^r)$, 
we finally obtain
\be
\hat \sigma(\chi,\delta) =  \left[  a_{\mathcal{H}}- \frac 8 3 \left(\tan \chi  \cos \delta  
- s_{\mathcal{H}}  \sqrt{\frac{1 + a_{\mathcal{H}}}{8}} \right)^2  \right] \cos^4 \chi \,.
\ee

Observe that $\hat \sigma$ depends on the vevs of moduli only through the quantity $a_{\mathcal H}$.\footnote{Certainly, 
for a given choice of the superpotential, $\chi$ and $\delta$ also depend on the moduli. Nevertheless, in the present approach 
$\chi$ and $\delta$ are independent of the moduli in the sense that we are leaving free the parameters entering the superpotential 
that {\it a posteriori} will do the job of stabilising the moduli. How to determine these parameters will be the subject of Section 
\ref{sec: de Sitter - two mod}.} 
Notice also that the squared term can always be set to zero by tuning $\chi$. On the other hand, as long as $\Delta <0$ 
the term proportional to $ a_{\mathcal{H}} $ is always positive. For a fixed value of $a_{\mathcal{H}} \in [0,+\infty)$, we 
may then compute the maximal value $\hat \sigma_0$ that can be achieved for $\hat \sigma$. This corresponds 
to finding the optimal direction $f_0^i$ discussed in Section \ref{Section: sGoldstino}. The relevant 
extremum occurs at
\be
\delta_0 = 0 \,, \qquad \tan \chi_0 = s_{\mathcal{H}} \sqrt{\frac {1+ a_{\mathcal{H}}}8}\,(1+\epsilon) \,,
\ee
where $\epsilon$ is a quantity still to be determined. One has then 
\be
\hat \sigma_0 = \frac {64\, \big[a_{\mathcal{H}} - (1 + a_{\mathcal{H}})\, \epsilon^2/3 \big]}{\big[8 + (1+a_{\mathcal{H}}) (1+\epsilon)^2\big]^2} \,.
\label{sigmahetepsilon}
\ee
Notice first that one gets a lower bound on the size that $\hat \sigma$ can reach by setting $\epsilon \simeq 0$,
which corresponds to setting to zero the negative definite part of the numerator. This is what was done in 
\cite{Covi:2008cn}, and results in the value $\hat \sigma_0 \simeq 64\, a_{\mathcal{H}}/(9+a_{\mathcal{H}})^2$. 
This expression has an extremum at $a_{\mathcal{H}} = 9$ where it reaches its maximal value $\hat\sigma_0 \simeq 16/9$. 
The true maximal value $\hat \sigma_0$ is however obtained for a non-vanishing value of $\epsilon$ determined by the 
stationarity condition $\partial \hat \sigma/\partial \epsilon = 0$, which is a cubic polynomial. This polynomial accidentally factorises 
in a simple way in this case, and it is actually possible to find the following simple expression for the value of $\epsilon$:
\be
\epsilon = \frac 32 \Big( \frac {\sqrt{1+ a_{\mathcal{H}}/9}}{\sqrt{1+a_{\mathcal{H}}}} - 1 \Big) \,.
\label{epsilonex}
\ee
Notice that $\epsilon$ is only small for small $a_{\mathcal{H}}$. This means that the exact $\hat \sigma_0$
will depart significantly from the approximate one for large values of $a_{\mathcal{H}}$. Plugging (\ref{epsilonex}) back into 
(\ref{sigmahetepsilon}) one finds that this is given by:
\be
\hat \sigma_0 = \frac{128}{3} \frac{ a_{\mathcal{H}} + 9 \sqrt{(1 + a_{\mathcal{H}})(1 + a_{\mathcal{H}}/9) } - 9}
{\left( 21 + a_{\mathcal{H}} - 3 \sqrt{(1 + a_{\mathcal{H}})(1 + a_{\mathcal{H}}/9) } \right) ^2  } \,,  \label{sigma-heterotic-a}
\ee
From eq.~(\ref{sigma-heterotic-a}) we see that $\hat \sigma_0$ grows asymptotically as 
$2/3 \,a_{\mathcal{H}}$ for large values of $a_{\mathcal H}$ and can thus be made arbitrarily 
large and positive. This means that for heterotic models the sGoldstino mass scale $m$ can be made 
arbitrarily large by tuning the value of the moduli. As we shall see in the following subsection, this is not the 
case for orientifold models with two moduli.

\subsection{Orientifold models}

Let us consider now the case of orientifold models. We focus on type IIB models with O3/O7 planes, 
where the effective K\"ahler potential in the large-volume limit takes the form \cite{Grimm:2004uq}
\be
K = - 2 \log \mathcal V \,, \qquad \mathcal V = \frac{1}{48} d^{ijk} v_i v_j v_k \,.   \label{eq:orientifoldkaehler}
\ee
In this expression $d^{ijk}$ denotes the collection of intersection numbers of the CY (rescaled by a factor 
of $1/8$ for convenience) and $v_i$ are the volume moduli. However, the $v^i$ do not directly correspond to the real part of
scalar components of chiral superfields in this case. These are instead given by new fields 
$\rho^i $, related to the $v_i$ via the quadratic relation
\be
\rho^i = \dfrac{\partial \mathcal V}{\partial v_i}= \frac{1}{16}\, d^{i j k} v_{j} v_{k} \,. 
\label{eq: var-change}
\ee
One then has to invert this relation and express the $v_i$ in terms of the $\rho^i$. After that, one obtains the superfield 
dependence of $K$ by setting $\rho^i = (T^i + \bar T^i)/2$. In general, this can however not be given explicitly and the K\"ahler potential (\ref{eq:orientifoldkaehler}) remains an implicit function of the $T^i$. Note finally that 
we have used lower indices for the fields $v_i$ in order to get upper indices for the fields $\rho^i$. Correspondingly 
we have used upper indices for the intersection numbers $d^{ijk}$, but it should be stressed that they are the same 
objects as in the heterotic case. 

From the above implicit definition of the K\"ahler potential it follows that $K_{i} = -\,\frac 12\, e^{K/2} v_i$ and $K^i = - (T^i + \bar T^i)$.
The metric and the Riemann tensor are then found to be (see \cite{Covi:2008ea} and \cite{DFT} for more details):
\bea
g_{i j} &=& K_i K_j + e^{-K} \dm_{ijk} K^k \,, \\
R_{i j m n} &=& - g_{im} g_{jn} + e^{-2K} \big(\dm_{i j k} g^{kl} \dm_{l m n} 
+ \dm_{i n k} g^{kl} \dm_{l j m} \big)  + g_{in} K_j K_m + g_{jm} K_i K_n  \nn\\
&\;&+\, g_{im} K_j K_n + g_{jn} K_i K_m + g_{ij} K_m K_n + g_{mn} K_i K_j - 3 K_i K_j K_m K_n \nn \\
&\;& -\, e^{-K} \big(\dm_{imj} K_n + \dm_{imn} K_j + \dm_{inj} K_m +
\dm_{nmj} K_i \big)  \,,
\label{Riemann-orientifold}
\eea
where we introduced the notation
\be
\dm_{ijk} \equiv g_{ip} g_{jq} g_{kl} d^{p q l} \,.
\ee
Inserting these expressions into the definition of $\hat \sigma(f^i)$ in (\ref{sigma}) and using the parametrisation 
(\ref{f-2-mod}) for $f^i$ we can as before rewrite $\hat \sigma(f^i)$ in the form 
$\hat \sigma (f^i) = - 2 \hat s^i \hat s_i + \hat \omega$ where:
\bea
\hat s^i &=& n^i \left[\frac{2}{\sqrt{3}}  \tan \chi \cos \delta - \frac 12 e^{-K} d^{p q r} n_p n_q n_r \right] \cos^2 \chi , \\
\hat \omega &=& \bigg[1 - \frac32 
\Big(e^{-K} d^{p q r} n_p n_q n_r \Big)^2 \bigg] \cos^4 \chi\,.
\eea
On the other hand, it can be shown that~\cite{Covi:2008ea}
\begin{eqnarray}
1 - \frac{3}{2} \Big(e^{-K} d^{p q r} n_p n_q n_r \Big)^2 =  a_{\mathcal{O}} \, ,
\end{eqnarray}
where
\be
a_{\mathcal{O}} \equiv   \frac{\Delta}{24} \, \frac{(\det g)^3}{e^{4 K}}   \le 1\,.
\ee
Putting all of this together, and introducing the sign $s_{\mathcal{O}} = \mathrm{sign}(d^{p q r} n_p n_q n_r)$,
we finally obtain
\be
\hat \sigma (\chi,\delta) =  \left[a_{\mathcal{O} }- \frac 83 
\left(\tan \chi \cos \delta - s_{\mathcal{O}}  \sqrt{\frac{1 - a_{\mathcal{O}}}{8}} \right)^2 \right] \cos^4 \chi \,.
\ee
It is clear that, as before, the squared term can always be set to zero by tuning $\chi$ and then $\hat\sigma>0$ 
as long as the term proportional to $a_{\mathcal{H}}$ is positive, which is the case when $\Delta>0$.

As in the previous subsection, we can now ask what is the maximum value for $\hat \sigma$ obtained 
by varying the Goldstino direction $f^i$, for a given $a_{\mathcal{O}} \in [0,1]$. 
The relevant extremum occurs for 
\be
\delta_0 = 0 \,, \qquad \tan \chi_0 = s_{\mathcal{O}} \sqrt{\frac {1- a_{\mathcal{O}}}8}\,(1+\epsilon) \,.
\ee
One then has
\be
\hat \sigma_0 = \frac {64\, \big[a_{\mathcal{O}} - (1-a_{\mathcal{O}})\, \epsilon^2/3\big]}
{\big[8 + (1-a_{\mathcal{O}}) (1+\epsilon)^2\big]^2} \,.
\label{sigmaorepsilon}
\ee
One gets as before a lower bound on $\hat \sigma_0$ by setting $\epsilon \simeq 0$.
This gives the approximate value $\hat \sigma_0 \simeq 64 \, a_{\mathcal{O}}/(9-a_{\mathcal{O}})^2$, which 
grows as $a_{\mathcal{O}}$ is increased until the point $a_{\mathcal{O}} = 1$, where it reaches its maximal 
value $\hat \sigma_0 \simeq 1$. But again the exact maximal value of $\hat \sigma$ for 
a given $a_{\mathcal{O}}$ is larger and occurs for a in general non-vanishing value of $\epsilon$ determined 
by the condition $\partial \hat \sigma/\partial \epsilon = 0$, which is again a cubic polynomial. In this 
case, this polynomial is generic, and the expression for the value of $\epsilon$ is somewhat complicated. 
One finds:
\be\label{ep}
\epsilon =  \frac {\sqrt{1+ 5\, a_{\mathcal{O}}/9}}{\sqrt{1-a_{\mathcal{O}}}} \big(3 \sin \theta - \sqrt{3} \cos \theta\big) \,,
\ee
where 
\be
\theta \equiv \frac{1}{3} \arccos \left(\frac {a_{\mathcal{O}}}{\sqrt{3}} \frac {\sqrt{1-a_{\mathcal{O}}}}
{(1 + 5\, a_{\mathcal{O}}/9)^{3/2}} \right)\,.
\ee
Plugging this back into (\ref{sigmaorepsilon}), one finds that the exact maximal value $\hat \sigma_0$
is given by a relatively complicated expression, which we do not report here. Fortunately, one can 
however check that the quantity $\epsilon$ given by (\ref{ep}) is always quite small for any value of 
$a_{\mathcal{O}} \in [0,1]$. In particular, one easily verifies that also the exact 
$\hat \sigma_0$ increases monotonically as a function of $a_{\mathcal{O}}$, and that for 
$a_{\mathcal{O}} = 1 $ one obtains $\hat \sigma_0 = 1$. 
In practise one can then approximate the maximal value of $\hat \sigma$ with the one associated 
with $\epsilon \simeq 0$, namely
\be
\hat \sigma_0 \simeq \frac {64\, a_{\mathcal{O}}}{(9-a_{\mathcal{O}})^2} \,.
\ee
Notice finally that the fact that $\hat \sigma$ can be at most $1$ 
implies the following upper bound for the sGoldstino mass scale $m$:
\be
m^2\leq (3 + \gamma) m_{3/2}^2\, .
\ee
This is an interesting result concerning the phenomenology of 
orientifold compactifications. It asserts that the lightest modulus 
cannot be much heavier than the gravitino. 
It seems therefore to point towards a large gravitino mass as the
only way to ease the cosmological moduli problem \cite{moduli}.
As we shall see during the next section, one can actually saturate the above bound by suitably 
tuning the superpotential.

\section{Constructing de Sitter vacua  with two moduli} \setcounter{equation}{0} \label{sec: de Sitter - two mod}

Let us now come to the main point of this paper, namely to the question of how for a given K\"ahler potential,
satisfying the necessary condition for metastability on the sign of $\Delta$, one may construct superpotentials 
that indeed allow for local minima of the scalar potential $V$ with a non-negative cosmological constant. 
Our strategy will be to assume some reference values for the fields at the location of the minimum, $T^{1,2}=T^{1,2}_{0}$,
and then to reconstruct the local behaviour that $W$ needs to have at that point.\footnote{One may also 
try to brutally scan over the parameter space of some plausible superpotential for those models 
that satisfy the metastability necessary condition. However, this proves to be very cumbersome as soon 
as there are several parameters. In this framework, the algebraic method for finding dS minima developed 
in ref.~\cite{Lukas} may perhaps be useful.}
We will thus consider an expansion of the form:
\bea
W(T) &=&  W_0 + W_i (T-T_0)^i + \frac 12 W_{i j} (T - T_0)^i (T-T_0)^j \nn\\[-1mm]
&\;& +\, \frac 16 W_{i j k} (T-T_0)^i (T-T_0)^j (T-T_0)^k  + \cdots \,. 
\label{W-expan}
\eea
The goal is to determine suitable coefficients $W_0$, $W_i$, $W_{i j}$ and  $W_{i j k}$. Higher order terms in the 
expansion do not affect the masses of scalar fluctuations around the vacuum and can therefore be omitted. 
Since we are demanding stabilisation at $T^{1,2}=T^{1,2}_{0}$, these coefficients depend on $T^{1,2}_{0}$ 
via $K$ and its derivatives evaluated at these field values. More precisely, they depend only on 
${\rm Re}\,T_{0}^{1,2}$, because of the shift symmetry of $K$. Hence, the vevs 
of the axions ${\rm Im}\, T^i$ do not affect the coefficients in eq.~(\ref{W-expan}) and can be chosen freely. 

Let us now describe a systematic procedure to reconstruct the coefficients $W_0$, $W_i$, $W_{i j}$ and  $W_{i j k}$.
Notice, before starting, that the freedom in choosing the two vevs $T_0^{1,2}$ can be used to achieve any desired
value for the volume $\mathcal{V}$, and a suitable positive value for the parameter $a$. More precisely, the value 
of $a$ fixes the ratio of $T_0^1$ and $T_0^2$, whereas the value of the volume $\mathcal{V}$ fixes their overall size.
Note also from eq.~(\ref{W-expan}) that rescaling the vevs of the fields $T^{1,2}_{0}$ can be compensated
by rescaling the coefficients appropriately, after factorising out the overall superpotential scale $W_0$.

\subsection{Tuning $W_0$}

The coefficient $W_0$ is fixed, modulo a phase that we shall discard, by the value one desires to achieve for the 
gravitino mass compared to the volume. From the definition of $m_{3/2}$ one gets the relation
\be
|W_{0}| = m_{3/2} \,e^{-K/2} \,.
\ee
Note that due to the different definitions of the volume $\mathcal{V}$ for heterotic and orientifold models, this 
equation translates into different relations between $m_{3/2}$ and $\mathcal{V}$ in heterotic and orientifold
models. In the two cases one finds respectively
\be
|W_0 | = m_{3/2} \mathcal{ V_{\cal H}}^{1/2}, \qquad | W_0 | = m_{3/2} \mathcal{ V_{\cal O}},
\ee
In any case, the value of $ W_0$ fixes the overall scale of the potential.

\subsection{Tuning $W_i$}

The two coefficients $W_i$ are fixed by the value of the cosmological constant and the 
direction of supersymmetry breaking that one desires to achieve. Indeed, one has by definition 
$G_i=K_i + W_i/W_0$, and $G_i$ can be parametrised in terms of $\gamma$ and $f_i$ as 
$G_i = \sqrt{3(1+\gamma)} f_i$. Recalling also the definition $K_i = \sqrt{3}\, k_i$, it follows then that:
\begin{eqnarray} \label{eq:lnW_i}
\frac{W_i}{W_0}= \sqrt{3} \left( \sqrt{1+\gamma} \, f_{i} - k_i \right) \,. 
\end{eqnarray}
This fixes $W_i/W_0$ in terms of $\gamma$ and $f_i$. The direction $f_i$, which we have parametrised 
by $\chi$ and $\delta$ in eq.~(\ref{f-2-mod}), must be chosen inside a cone sufficiently close to the optimal 
direction $f_{0i}$, in such a way that $m^2 > 0$.\footnote{Note that in eq.~(\ref{eq:lnW_i}) the overall 
phase discarded in the parametrisation (\ref{f-2-mod}) becomes relevant and represents
an additional parameter that one can tune.}

\subsection{Tuning $W_{ij}$} \label{stationary}

The three coefficients $W_{ij}$ are fixed by demanding stationarity of the potential, $\nabla_i V=0$, and 
positivity of the two-dimensional diagonal blocks $V_{i \bar \jmath}$ of the mass matrix, which is necessary
for positivity of the full mass matrix. It is convenient to first implement the stationarity conditions (\ref{stationary1}).
This implies the following two relations, which allow to fix two of the three parameters $W_{ij}$ in terms of the last 
one (understanding now $G_i$ as fixed):
\be
\label{eq:condition_Wii}
\frac{W_{ij}}{W_0}  G^j = -(1 + 3 \gamma)  G_i -  G_{\bar \imath} 
+ \Gamma_{ij}^{k} G_k  G^j + \frac{W_i W_j}{W_0^2}  G^j \,.
\ee
The remaining parameter among the $W_{ij}$ which is still free is then fixed by demanding positivity of 
the two-dimensional matrix $V_{i \bar \jmath}$. This amounts to requiring that its two eigenvalues 
are positive. Notice that we have already ensured the positivity of the projection $m^2 = V_{i \bar \jmath} f^i f^{\bar \jmath}$. Thus, it makes sense now to study the projection of $V_{i \bar \jmath}$ along the remaining direction $u^i$ orthogonal to $f^i$ in order to understand when the positivity of the whole matrix $V_{i \bar \jmath}$ is possible. 
This direction is completely fixed, again modulo an overall phase that we do not display, and is given by: 
\bea
u^i &=& \cos \chi\, k^i -  e^{i \delta} \sin \chi\, n^i \,, \qquad
u_i = \cos \chi\, k_i -  e^{-i \delta} \sin \chi \, n_i \,, \nn \\
u^{\bar \imath} &=& \cos \chi\, k^i -  e^{-i \delta} \sin \chi\, n^i \,, \hspace{17pt}
u_{\bar \imath} = \cos \chi\, k_i -  e^{i \delta} \sin \chi\, n_i \,.  \label{u-2-mod}
\eea
We are then led to compute
\be
m{'}^{2} \equiv V_{i \bar \jmath} \, u^i u^{\bar \jmath} \,.
\ee
Using the fact that $\nabla_i G_{j} u^i f^j = 0$ by the stationarity condition, one finds that this 
second mass scale is given by:
\be
m{'}^{2} =  \Big[1 + 3 \gamma - 3 (1+\gamma) \hat \beta (u^i) + |\nabla_i G_j u^i u^j |^2 \Big] m_{3/2}^2 \,, 
\label{lambda-2}
\ee
where 
\be
\hat \beta(u^i) = R_{i j m n} \,  u^i  u^{\bar \jmath}  f^m f^{\bar n} \,.
\ee
From eq.~(\ref{lambda-2}) we see that it is always possible to tune the quantity $\nabla_i G_j$ in order
to make the last positive term dominate and achieve $m{'}^2 > 0$, compatibly with the 
two stationarity conditions that also involve $\nabla_i G_j$, since there are three parameters $W_{ij}$.
On the other hand, the matrix $V_{i \bar \jmath}$ has in general a non-zero mixing between the $f^i$
and $u^i$ directions, which is given by 
\be
V_{i \bar \jmath} \, u^{i} f^{\bar \jmath} = - 3 (1+\gamma) m_{3/2}^2 \, R_{i j m n}  u^i  f^{\bar \jmath}  f^m f^{\bar n}. \label{off-diag-i-jbar}
\ee
Since this quantity is independent of $\nabla_i G_j$, it is now evident that it is always possible to tune the value of $m'^2$ until both eigenvalues of $V_{i \bar \jmath}$ become positive.

A simple although not mandatory possibility to fix unambiguously the free parameter left among the 
$W_{ij}$ after imposing the stationarity condition is to require that $f_i$ should be aligned along the 
optimal direction $f_0^i$ maximising $m^2$. In that case the orthogonal direction $u_i$ is then also 
fixed to some $u_0^i$. In this situation, eq.~(\ref{f-maximal}) implies that one has
$V_{i \bar \jmath} u_0^i f_0^{\bar \jmath} = 0$, so that $m^2$ and $m'^2$ coincide with the two 
eigenvalues of $V_{i \bar \jmath}$. Additionally, the quantity $\hat \beta$ takes a 
definite value, which is different for heterotic and orientifold models and depends on $a_{\mathcal{H}}$ 
and $ a_{\mathcal{O}}$ respectively. After a straightforward but lengthy computation one finds:
\bea
\hat \beta_{\mathcal{H} 0} &=& \frac 1{24} \Big(9 - 2 a_{\mathcal{H}} + (7 + 2 a_{\mathcal{H}}) \cos 4 \chi_0 
+ 4 s_{\mathcal{H}} \sqrt{2 (1+a_{\mathcal{H}})} \sin 4 \chi_0 \Big) \,, \\
\hat \beta_{\mathcal{O} 0} &=& \frac 1{24} \Big(9 - 4 a_{\mathcal{O}} + (7 + 4 a_{\mathcal{O}}) \cos 4 \chi_0  
+ 4 s_{\mathcal{O}} \sqrt{2(1-a_{\mathcal{O}})} \sin 4 \chi_0 \Big) \,.
\eea
In these expressions, the quantity $\chi_0$ is the one that leads to the maximal value $\hat \sigma_0$ 
for $\hat \sigma$, namely $\tan \chi_0 = s \sqrt{(1 \pm a)/8} (1 + \epsilon)$. 
For heterotic models, one has to use the exact value (\ref{epsilonex}), but for orientifold it is good enough to use the approximate value 
$\epsilon \simeq 0$. In this way one finds:
\bea
\hat \beta_{\mathcal{H} 0} &=&\frac {9 - a_{\mathcal H} + 9 \sqrt{(1 + a_{\mathcal H})(1 + a_{\mathcal H}/9)}}{27 + 2 a_{\mathcal H}} \,, \\
\hat \beta_{\mathcal{O} 0} &\simeq& \frac 23 \bigg(1 - 12 \frac {a_{\mathcal O} (1 - a_{\mathcal O})}{(9 - a_{\mathcal O})^2} \bigg)\,.
\eea
We see in particular that both quantities remain bounded respectively by $1$ and $2/3$ in the allowed ranges for $a$.

\subsection{Tuning $W_{ijk}$} \label{off-diagonal-stable}

Finally, the four coefficients $W_{ijk}$ need to be chosen in such a way that all of the four eigenvalues of the full mass matrix 
$M^2$ are positive, even after taking into account the effect of the off-diagonal block $V_{ij}$. Solving then 
the expression for $V_{ij} $ in terms of the $W_{ijk}$, one deduces the following three relations (where now 
both $G_i$ and $\nabla_i G_j$ are understood as fixed):
\begin{eqnarray}
\frac{W_{i j k}}{W_0}  G^k &=& 
\bigg[ R_{i j k m} G^{\bar m} + \Gamma_{i j}^{m} \nabla_m G_k + \Gamma_{(i k}^{m} \nabla_m G_{j)}  
- 2 \frac{W_i W_j W_k}{W_0^3} + 2 \frac{W_{(i} W_{j) k}}{W_0^2} + \frac{W_k W_{i j}}{W_0^2} \nonumber \\  
&\;& \;+\, \Gamma_{(i k}^{m} \bigg(\frac{W_{m j)}}{W_0} - \frac{W_m W_{j)}}{W_0^2}\bigg) \bigg]  G^k    
- (2 + 3 \gamma) \nabla_{(i} G_{j)} + 3 \gamma \, G_i G_j + \frac{V_{i j} }{m_{3/2}^2}\,.
\label{eq:condition_Wiii}
\end{eqnarray}
Recall that for $V_{ij}=0$ the mass spectrum is degenerate, with two states for each of the two eigenvalues of $V_{i \bar \jmath}$,
which have already been adjusted to be positive with the previous step. When instead $V_{ij} \neq 0$, the spectrum splits
and one has to make sure that no eigenvalue becomes negative. This represents three constraints
on the four parameters  $W_{ijk}$. If for simplicity one requires $V_{ij}=0$, then these become three relations, which allow to 
express three of the four parameters  $W_{ijk}$ in terms of the last one. More generally, we can leave $V_{i j} $ arbitrary and 
compute the four eigenvalues as functions of the $W_{i j k}$'s. In generic situations it is hard to do this in an analytic way, 
but it can be easily done with computer assistance. One can then scan this multi-parameter space for regions where all 
masses are positive. 

\vskip 15pt

The next step is to match these `local superpotentials' with the expansion of some string-motivated superpotential around the 
given vevs. To this end we will consider in the next section superpotentials with enough parameters and determine these 
parameters in such a way that the Taylor expansion around the extremum matches the cubic superpotential constructed as 
outlined above.
 
\section{Examples of models with dS vacua} \setcounter{equation}{0}  \label{sec: string models}

Let us now apply the procedure described in last section to construct some illustrative examples of string models 
with a sector of two volume moduli admitting a metastable dS vacuum. For simplicity, we shall focus on the case 
where the cosmological constant vanishes ($\gamma=0$) and on separable superpotentials of the form 
$W(T^1,T^2) = W^{(1)}(T^1) + W^{(2)}(T^2)$. This choice implies further restrictions on the coefficients of the Taylor 
expansion of the superpotential about the vacuum, namely $W_{12} = W_{112} = W_{221} = 0$, and the existence 
of a solution with these characteristics is no longer guaranteed from the beginning. We will however see that it is 
nevertheless possible to find simple examples of this type.

\subsection{Orientifold models}

Let us start with orientifold models. For these models, the way in which the dilaton and the complex structure moduli
may be stabilised is well understood \cite{Giddings:2001yu}, and restricting to the sector of volume moduli may be 
justified. In this case, the necessary condition for metastability is that the discriminant $\Delta$ should be positive. 
As a prototype example, let us take a CY manifold with intersection numbers given by
$ d^{111} = -1$,  $d^{112}=0$, $d^{122} = 1$ and $d^{222}=0$, for which $\Delta= 108>0$. The
K\"ahler potential is then found to take the following form:
\bea
K &=& - \log \bigg[\frac 89 \Big((T^1 \hspace{-2.5pt}+\hspace{-1pt} \bar T^1) 
+\! \textstyle{\sqrt{(T^1 \hspace{-2.5pt}+\hspace{-1pt}  \bar T^1)^2 \!+ (T^2 \hspace{-2.5pt}+\hspace{-1pt} \bar T^2)^2}}\Big) \nn \\
&\;& \hspace{30pt} \bigg(\frac {(T^2 \hspace{-2.5pt}+\hspace{-1pt} \bar T^2)^2 
\!+(T^1 \hspace{-2.5pt}+\hspace{-1pt} \bar T^1)^2 \!- (T^1 \hspace{-2.5pt}+\hspace{-1pt} \bar T^1) 
\textstyle{\sqrt{(T^1\hspace{-2.5pt}+\hspace{-1pt} \bar T^1)^2 \!+ (T^2 \hspace{-2.5pt}+\hspace{-1pt} \bar T^2)^2}}}
{T^2 \hspace{-2.5pt}+\hspace{-1pt}  \bar T^2} \bigg)^2\bigg] \,.
\eea
We require that at the stationary point one should have $a_{\mathcal{O}}=1$. As seen in Section 3.4, this choice
allows to maximise the sGoldstino mass and corresponds to setting $\hat s^i = 0$. We will moreover require that the 
volume takes some definite numerical value $\mathcal{V}_{\cal O}$. These two conditions fix the vevs of the two fields 
to the following values, in units of $ \mathcal{V}_{\cal O}^{2/3}$: 

\begin{table}[h]
\vskip 5pt
\begin{displaymath}
\begin{array}{|l|l|}
\hline 
\; T_0^1 \;&\; \;\;\; 0.412741 \; \raisebox{14pt}{$$}\\
\; T_0^2 \;&\; \;\;\; 0.714888  \raisebox{-7pt}{$$}\\
\hline
\end{array}
\end{displaymath}
\vskip -40 pt
\be
\ee
\vskip 10pt
\end{table}

Applying then the procedure described in the previous section, in such a way to achieve some definite numerical value $m_{3/2}$ 
for the gravitino mass, we find that the local behaviour that the superpotential needs to have is specified by the following 
Taylor coefficients, in units of $m_{3/2} \mathcal{V}_{\cal O}$ for $W_0$, $m_{3/2} \mathcal{V}_{\cal O}^{1/3}$ for $W_{i}$,
$m_{3/2} \mathcal{V}_{\cal O}^{-1/3}$ for $W_{ii}$ and $m_{3/2} \mathcal{V}_{\cal O}^{-1}$ for $W_{iii}$:
\begin{table}[h]
\begin{displaymath}
\begin{array}{|l|l|}
\hline 
\;W_{0}  \;&\; \;\;\; 1.000000 \; \raisebox{14pt}{$$} \\
\;W_{1}  \;&\; \;\;\; 2.021311 \\
\;W_{2}  \;&\; \;\;\; 0.931223  \\
\;W_{11}  \;&\; \;\;\; 0.999657\\
\;W_{22}  \;&\; -0.797685 \\
\;W_{111}  \;&\; -0.827204 \\
\;W_{222}  \;&\; \;\;\; 3.308820  \raisebox{-7pt}{$$} \\
\hline
\end{array}
\end{displaymath}
\vskip -77 pt
\be
\label{localo}
\ee
\vskip 40pt
\end{table}

\noindent
In this way, the four physical square-mass eigenvalues $m_i^2$ at the minimum, obtained after canonically 
normalising the fields, are given by $2.77$, $2.95$, $3.86$, $5.14$ in units of $m_{3/2}^2$.

Notice that the coefficients (\ref{localo}) scale in the following way with the size $T_0 \sim \mathcal{V}_{\cal O}^{2/3}$
of the field vevs:
\be \label{eq:hierarchy}
W_0 : W_i : W_{i i}:W_{i i i} \sim 1 : T_0^{-1} : T_0^{-2} : T_0^{-3}\, .
\ee
This scaling can be understood as naturally following from the structure of eqs.~(\ref{eq:lnW_i}), 
(\ref{eq:condition_Wii}) and (\ref{eq:condition_Wiii}), although it is conceivable that it could be changed 
with some additional fine-tuning of the parameters of the theory. This relation calls nevertheless for superpotentials
with derivatives satisfying $T^n W^{(n)}/W \sim 1$. 

Let us now try to match the coefficients (\ref{localo}) of the local expansion with an explicit superpotential of a form 
that may plausibly arise in these models. The simplest possibility is to try with an exponential effective superpotential that typically 
arises from gaugino condensation. This has the simple form $W = A e^{-a T}$, provided that $a T \gg 1$, corresponding
to a weekly coupled four-dimensional low-energy effective theory. For this type of superpotential, however,
one gets $T^n W^{(n)}/W \sim (a T)^n$, which is much larger than $1$ as soon as $a T \gg 1$. It is then not 
possible to reproduce the scaling (\ref{eq:hierarchy}). This problem can however be cured by adding a constant
term $W = \Lambda$, or possibly also a linear term $W = F T$, which may for instance arise from background 
fluxes.\footnote{This kind of effect has also been used to construct supersymmetric vacua. See for instance 
refs.~\cite{micu,Palti:2007pm}.} 
Notice finally that one needs a superpotential with at least 7 free parameters in order to be able to match all the local 
coefficients. 

As a simple and `symmetric' possibility to try out, one could consider a superpotential with a constant term plus a 
racetrack term for each field:
\be\label{eq:Wrt2}
W = \Lambda + A_1 e^{- a_1 T^1}+ A_2 e^{- a_2 T^2}+ B_1 e^{- b_1 T^1}+ B_2 e^{- b_2 T^2} \,.
\ee
Such a combination of exponentials could arise for instance from gaugino condensation on two sets of D7-branes 
wrapping cycles controlled by the moduli $T^1$ and $T^2$, each giving rise to a gauge group consisting of two semisimple factors.
This $W$ has 9 coefficients which have to satisfy 7 equations. This allows to express 7 of them in terms of the other 2, 
say $b_1$ and $ b_2 $, and of the coefficients of the local superpotential. Among other relations, one finds that
\be \label{eq:ai}
a_i=- \frac{b_i W_{i i} + W_{i i i}}{b_i W_i + W_{i i}} \,.
\ee
One can then choose the values of $b_i$ in such a way that $b_i T_0^i \gg 1$, but by eq.~(\ref{eq:hierarchy}) one 
will then get $a_i T_0^i  \sim 1 $. This means that the constant term allows to make only some of the exponents in 
the exponential terms large, and some of them remain of order one, so that higher-power corrections may become
relevant. An example of this type is obtained with the following values of the parameters, in units of 
$m_{3/2} \mathcal{V}_{\cal O}$ for $\Lambda,A_i,B_i$ and $\mathcal{V}_{\cal O}^{-2/3} $ for $a_i$,$b_i$:

\begin{table}[h]
\begin{displaymath}
\hspace{-124pt}
\begin{array}{|l|l|}
\hline 
\;\Lambda \;&\; \;\;\; 2.63036 \times 10^1 \; \raisebox{14pt}{$$} \\
\;A_1 \;&\; \;\;\; 7.37726  \times 10^1\\
\;B_1 \;&\; -9.77287 \times 10^1\\
\;A_2 \;&\; -1.50213 \times 10^0\\
\;B_2 \;&\; -2.80545 \times 10^0 \raisebox{-7pt}{$$} \\
\hline
\end{array}
\vspace{-93pt}
\end{displaymath}
\begin{displaymath}
\hspace{124pt}
\begin{array}{|l|l|}
\hline 
\;a_1 \;&\; \;\;\; 3.49830 \times 10^{-1} \; \raisebox{14pt}{$$} \\
\;b_1 \;&\; \;\;\; 2.79764 \times 10^{-1}\\
\;a_2 \;&\; \;\;\; 7.30908 \times 10^{0} \\
\;b_2 \;&\; \;\;\; 4.19646 \times 10^{-1} \raisebox{-7pt}{$$} \\
\hline
\end{array}
\end{displaymath}
\vskip -47pt
\be
\label{rt-fit}
\ee
\vskip 35pt
\end{table}

A more satisfactory but slightly more complicated model may be obtained by adding linear terms. 
Let us consider for example the following form of the superpotential:
\be \label{eq:Wlinear-double-rt}
W = \Lambda + F_1 T^1 + F_2 T^2 + A_1 e^{-a_1  T^1} + A_2 e^{-a_2 T^2}+B_1 e^{-b_1  T^1} + B_2 e^{-b_2 T^2} \,.
\ee
While one still has $ W_{iii}/W_{ii} =-a_i $, as this condition is unaffected by the addition of a linear term, 
the relation between the coefficients $ a_i, b_i $ and $ W_{iii}/W_{ii} $ gets now more complicated and less
constraining. This allows to find parameters such that all the exponents in the exponential terms are large.
A working example of this type is obtained with the following choice of parameters, in units of $m_{3/2} \mathcal{V}_{\cal O}$ 
for $\Lambda,A_i,B_i$, $m_{3/2} \mathcal{V}_{\cal O}^{1/3}$ for $F_i$ and $\mathcal{V}_{\cal O}^{-2/3} $ for $a_i,b_i$:

\begin{table}[h] 
\begin{displaymath}
\hspace{-244pt}
\begin{array}{|l|l|}
\hline 
\; \Lambda \;&\; -4.83093 \times 10^{-1} \; \raisebox{14pt}{$$} \\
\; A_1  \;&\; \;\;\; 5.14986 \times 10^{9} \\
\; B_1  \;&\; -1.55366 \times 10^{10} \\
\; A_2  \;&\; -4.16798 \times 10^{8}\\
\; B_2  \;&\; \;\;\; 2.38480 \times 10^{10} \raisebox{-7pt}{$$} \\
\hline
\end{array}
\vspace{-93pt}
\end{displaymath}
\begin{displaymath}
\hspace{4pt}
\begin{array}{|l|l|}
\hline 
\; a_1  \;&\; \;\;\; 6.69463 \times 10^{1} \; \raisebox{14pt}{$$} \\ 
\; b_1 \;&\; \;\;\; 6.99410 \times 10^{1}\\
\; a_2  \;&\; \;\;\; 3.55839 \times 10^{1}\\
\; b_2  \;&\; \;\;\; 4.19646 \times 10^{1} \raisebox{-7pt}{$$} \\
\hline
\end{array}
\vspace{-77pt}
\end{displaymath}
\begin{displaymath}
\hspace{251pt}
\begin{array}{|l|l|}
\hline 
\; F_1  \;&\; \;\;\; 2.05036 \times 10^0 \raisebox{14pt}{$$} \\
\; F_2  \;&\; \;\;\; 8.92014 \times 10^{-1} \; \raisebox{-7pt}{$$}  \\
\hline
\end{array}
\end{displaymath}
\vskip -13pt
\be \label{numbers-lin-fit}
\ee
\vskip 30pt
\end{table}

Note that in order to achieve large values of the exponents $a_i T_0^i$, $b_i T_0^i$ 
at the minimum in this kind of models, one necessarily needs a hierarchy 
between the coefficients $A_i$, $B_i$ of the gaugino condensation terms and 
the coefficients $\Lambda$ and (if present) $F_i$. Indeed, in order for all the 
terms in $W$ to be of comparable size at the minimum, the 
ratio of these two kinds of coefficients must be of order $e^{a_i T_0^i}$, $e^{b_i T_0^i}$.
In (\ref{rt-fit}) such a hierarchy is absent, because the exponents are of order one, 
whereas in (\ref{numbers-lin-fit}) it is large, because the exponents are large.

The particular numbers chosen in the second example serve as an illustration 
but can correspond to realistic values for physical parameters. 
The values $a_i T_0^i \sim b_i T_0^i \sim 25$ corresponds to the size of the MSSM
inverse couplings at the unification scale. Moreover, for a Weak scale gravitino 
mass $m_{3/2} \sim 10^{-16} M_{\rm Pl} \sim 100\, \mbox{GeV}$ and a reasonably 
large volume in Planck units $\mathcal V_{\mathcal O} \sim 10^3$, one has 
$A_i^{1/3}, B_i^{1/3} \!\! \sim 10^{-1} M_{\rm Pl} \sim 10^{17}~\mbox{GeV}$, which 
is a plausible gaugino condensation scale, and $\Lambda^{1/3} \sim 10^{-4} 
M_{\rm Pl} \sim 10^{14}\,\mbox{GeV}$, which could also be reasonable.

\subsection{Heterotic models}

Let us now consider heterotic models. In this case, the way in which the dilaton and the complex structure moduli 
may be stabilised is less understood, but we will nevertheless assume that these do not play any role and focus on two 
volume moduli. As an explicit example satisfying the necessary condition $\Delta < 0$, let us consider a 
CY manifold with intersection numbers $d_{111} = 1$, $d_{112}=0$, $d_{122} = 1$ and $d_{222}=0$,
for which $\Delta = - 108 < 0$. The corresponding K\"ahler potential is:
\be
K = - \log \Big[\frac 16 (T^1\!+ \bar T^1)^3 + \frac 12 (T^1\!+ \bar T^1) (T^2\!+ \bar T^2)^2\Big] \,.
\ee
We chose in this case the values of the field vevs in such a way that $a_{\mathcal{H}} = 9$, corresponding to setting $\hat s^i=0$. 
This choice does not correspond to the largest possible sGoldstino mass in this case, but it has the virtue of maintaining some 
similarity with the orientifold examples. Moreover, we require as before some definite numerical value $\mathcal{V}_{\cal H}$ for the volume.
This leads then to the following values of the vevs, in units of $\mathcal{V}_{\cal H}^{1/3}$:

\begin{table}[h]
\vskip 5pt
\begin{displaymath}
\begin{array}{|l|l|}
\hline 
\; T_0^1 \;&\; \;\;\; 0.405666 \; \raisebox{14pt}{$$}\\
\; T_0^2 \;&\; \;\;\; 0.749277 \raisebox{-7pt}{$$}\\
\hline
\end{array}
\end{displaymath}
\vskip -40 pt
\be
\ee
\vskip 10pt
\end{table}

\noindent
Applying the procedure outlined in the previous section, one finds the following set of local parameters, in 
units of $m_{3/2} \mathcal{V}_{\cal H}^{1/2}$ for $W_0$, 
$m_{3/2} \mathcal{V}_{\cal H}^{1/6}$ for $W_{i}$, $m_{3/2} \mathcal{V}_{\cal H}^{-1/6}$ for $W_{ii}$ and 
$m_{3/2} \mathcal{V}_{\cal H}^{-1/2}$ for $W_{iii}$:

\begin{table}[h]
\vskip 45 pt
\be
\label{lochet}
\ee
\vskip -85pt
\begin{displaymath}
\begin{array}{|l|l|}
\hline 
\;W_{0} \;&\; \;\;\; 1.00000 \; \raisebox{14pt}{$$} \\
\;W_{1} \;&\; \;\;\; 1.64415\\
\;W_{2} \;&\; \;\;\; 2.60392 \\
\;W_{11} \;&\; -17.4400 \\
\;W_{22} \;&\; \;\;\; 3.82418 \\
\;W_{111} \;&\; \;\;\; 616.732 \\
\;W_{222} \;&\; \;\;\; 2.31275 \raisebox{-7pt}{$$} \\
\hline
\end{array}
\end{displaymath}
\vskip-5pt
\end{table}

\noindent
In this model, the four physical square-mass eigenvalues $m_i^2$ at the minimum are given by $4.43$, $5.95$, $203.88$ 
and $311.92$ in units of $m_{3/2}^2$.

We may now proceed as for orientifold models and fit these coefficients with a superpotential involving exponential, 
constant or linear terms. In this case, however, the possible origin of such terms is less clear than for orientifolds. 
For instance, gaugino condensation produces exponential contributions, but with an exponent involving in first 
approximation only the dilaton. It is however common that the effective gauge coupling receives perturbative 
threshold corrections depending on the volume moduli as well. Assuming then that the dilaton does not play any 
role and the volume moduli are large, one can be left with an exponent linear in $T$. 
Notice moreover that taking this perspective there is no reason to require 
any longer that the exponent should be large and positive (see for example \cite{serone,Badziak:2008gv}). 
As a toy illustrative example with enough parameters, we can thus again consider a superpotential of the form (\ref{eq:Wrt2}). 
One can then, for example, reproduce the local coefficients (\ref{lochet}) with the following values of parameters, 
in units of $m_{3/2} \mathcal{V}_{\cal H}^{1/2}$ for $\Lambda,A_i,B_i$ and $\mathcal{V}_{\cal H}^{-1/3}$ for $a_i,b_i$:

\begin{table}[h]
\begin{displaymath}
\hspace{-124pt}
\begin{array}{|l|l|}
\hline 
\;\Lambda \;&\; -5.97604 \times 10^{-1} \; \raisebox{14pt}{$$} \\
\;A_1 \;&\; - 3.62358 \times 10^5\\
\;B_1 \;&\; -1.46692 \times 10^0\\
\;A_2 \;&\; \;\;\; 7.98841 \times 10^{-1}\\
\;B_2 \;&\; \;\;\; 7.49672 \times 10^{-1} \raisebox{-7pt}{$$} \\
\hline
\end{array}
\vspace{-93pt}
\end{displaymath}
\begin{displaymath}
\hspace{124pt}
\begin{array}{|l|l|}
\hline 
\;a_1 \;&\; \;\;\; 4.36876 \times 10^{1} \; \raisebox{14pt}{$$} \\
\;b_1 \;&\; \;\;\; 2.66924 \times 10^{0}\\
\;a_2 \;&\; -1.28225 \times 10^{0}\\
\;b_2 \;&\; \;\;\; 5.33848 \times 10^{0} \raisebox{-7pt}{$$} \\
\hline
\end{array}
\end{displaymath}
\vskip -50pt
\be
\ee
\vskip 35pt
\end{table}

As before, the hierachy arising between some of the coefficients $A_i,B_i$ and $\Lambda$
is related to the fact that some of the exponents $a_i T_0^i$, $b_i T_0^i$ are large 
at the minimum. In this case, for $m_{3/2} \sim 10^{-16}$ and 
$\mathcal V_{\mathcal H}\sim 10^3$ in Planck units, the particular numbers chosen 
in the example yield $A_i^{1/3}, B_i^{1/3} \sim 10^{13}-10^{15}~\mbox{GeV}$ and 
$\Lambda^{1/3} \sim 10^{13}\mbox{GeV}$.

\section{Conclusions} \setcounter{equation}{0}  \label{sec: conc}

In this paper we have developed a systematic method for constructing metastable dS vacua in 
supergravity models describing the volume moduli sector of CY string compactifications, 
without invoking subleading corrections breaking the no-scale property or uplifting terms. 
To do so, we have exploited the fact that there exists a necessary condition for the existence of 
metastable vacua, which constrains the allowed scalar geometry and supersymmetry-breaking 
directions \cite{Covi:2008ea}. We have focused on the simplest non-trivial case of two 
volume moduli, which allows for a detailed analysis, but we believe that the more complicated 
cases with more than two volume moduli can be treated similarly. 
We have singled out the special Goldstino direction which allows to maximise the moduli masses, 
and in the case of orientifold compactifications, we have found a strong upper bound of the lightest
modulus mass as a function of the gravitino mass.

The main result of the paper is an explicit procedure allowing to construct the local form of 
the superpotential that gives a metastable dS vacuum in models where the K\"ahler potential 
satisfies the necessary condition for metastability on the sign of the discriminant $\Delta$ of the 
intersection numbers $d_{ijk}$. 
We have also applied this procedure to construct a few simple examples of concrete models admitting 
viable metastable vacua that may plausibly emerge within heterotic and orientifold string compactifications 
with background fluxes and gaugino condensation effects. The fact that these models need to have
more than one dynamical field and at least seven independent parameters in the superpotential to 
allow for the construction is probably the reason why such models have not been noted earlier.
It is still an open question to study more realistic, more generic or even more minimal models, but 
we have now a proof of existence for dS vacua arising from simple F-term supersymmetry breaking 
in both the orientifold and heterotic case. It is also clear that the presence of vector multiplets giving a D-term contribution 
to supersymmetry breaking can potentially further improve the situation. More 
precisely, for a fixed value of $V$, increasing the ratio between the $D$-term and 
$F$-term contributions has the net effect of making the left-hand side of (2.11) 
smaller and therefore making that constraint milder, although the variety of allowed 
superpotentials is then reduced by the requirement of gauge invariance 
\cite{GomezReino:2007qi}. This helps, and in fact there exist no-scale models 
with a single chiral multiplet and a vector multiplet that admit metastable dS 
vacua \cite{Burgess:2003ic,carlos,Cremades:2007ig}.

We believe that our results emphasise in a clear way that it is actually possible to achieve genuine
metastable dS vacua even in models satisfying the no-scale property, provided that the scalar 
geometry is sufficiently generic. This is the case for the volume moduli sector of smooth CY 
compactifications, as opposed to their orbifolds limits, when at least two moduli arise. 
But of course in order to construct a realistic model, there are several other issues to be addressed.
One of them is the detailed mechanism stabilizing the other moduli and the impact of their dynamics
onto the dS vacuum admitted by the volume moduli sector. Another is the life-time of the dS vacuum against 
decay to other supersymmetric AdS vacua that generically arise at different values of the fields
\cite{Coleman:1980aw,Weinberg:1982id,Banks:2002nm}. 

\section*{Acknowledgements}

This work was partly supported  by the German Science Foundation (DFG)
under the Collaborative Research Centre (SFB) 676, the Swiss National 
Science Foundation, and by the Netherlands Organisation for Scienti�c Research (NWO) under a VIDI and a VICI Innovative
Research Incentive Grant. M.~G.-R. and C.~G. are grateful to the Institute for Theoretical Physics 
of EPFL for hospitality during the completion of this work. G.~A.~P. would like to thank A. Ach\'ucarro for useful discussions.

\end{document}